\documentclass[aps,prd,amsmath,longbibliography]{revtex4-1}

\usepackage{amsmath}
\usepackage{float}
\usepackage{subcaption}
\usepackage{array}
\newcolumntype{P}[1]{>{\centering\arraybackslash}p{#1}}
\usepackage{graphicx}
\usepackage{amssymb,color}
\renewcommand{\bar}[1]{\overline{#1}}
\renewcommand{\bar}[1]{\overline{#1}}
\newcommand{\half}{{\frac{1}{2}}}

\def\e{\epsilon}
\def\Dslash{\raise.15ex\hbox{/}\kern-.7em D}
\def\Pslash{\raise.15ex\hbox{/}\kern-.7em P}

  \def\Cv{{\cal V}}

\def\t{\tau}

\newcommand{\ra}{\rangle}

\newcommand{\ben}{\begin{displaymath}}
\newcommand{\een}{\end{displaymath}}
\newcommand{\be}{\begin{equation}}
\newcommand{\ee}{\end{equation}}
\newcommand{\bea}{\begin{eqnarray}}
\newcommand{\eea}{\end{eqnarray}}

\newcommand{\eq}[1]{Eq.~(\ref{#1})}

\newcommand{\bfp}{{\bf p}}
\newcommand{\bfP}{{\bf P}}

\def\m{\mu}

\newcommand{\beqn}{\begin{equation}}
\newcommand{\eeqn}{\end{equation}}

\def\Pm{P_{\rm miss}}
 \def\D{\Delta}

\begin{document}

\title{
Light-Front Holography Model of the EMC Effect }

\author{Dmitriy N. Kim and Gerald A. Miller}

\affiliation{ 
Department of Physics,
University of Washington, Seattle, WA 98195-1560, USA}

\date{\today}

\begin{abstract}
A new two-component model of the EMC effect based on Light-Front Holographic QCD (LFHQCD) is presented. The model suggests the EMC effect is the result of the nuclear potential breaking SU(6) symmetry. The model separates the $F_2^A$ nuclear structure function into two parts: a free contribution, involving the addition of proton and neutron structure functions weighted by the number of protons and neutrons respectively, and a nuclear/medium modified contribution that involves nucleus independent universal function. Further, the model displays a correlation between the size of the EMC effect and the SRC pair density, $a_2$ - extracted from kinematic plateaus at around $x > 1$ in inclusive quasi-elastic (QE) scattering.

\end{abstract}  
\maketitle     

\section{Introduction}
Deep inelastic lepton-nucleus scattering experiments, involving squares of four momentum transfers ($Q^2$) between 10 and hundreds of GeV$^2$,  have shown that nuclear structure functions (per nucleon) are different than that of free nucleons. This phenomenon is known as the EMC effect, named after the European Muon Collaboration where it was first observed; see {\it e.g.} the review~\cite{Hen:2016kwk} and the original work~\cite{Aubert:1983xm,Gomez:1993ri}. This was a shocking result as it was assumed that deep inelastic scattering (DIS) off a nucleus $A$ was the same as scattering off $A$ nucleons. This experimental observation taught us quark interactions in nuclei are important, and that parton distribution functions (PDFs) depend on the nuclear environment.  \\

The EMC effect is not large, of order 10-15\%, but is of fundamental interest because it involves the influence of nuclear properties on scales that resolve the nucleon size. However, scales larger than the nucleon size are relevant because modifications of nucleon structure must be caused by interactions with nearby nucleons. Indeed, after the nucleon size, the next largest length is the inter-nucleon separation length, $d$; this is the scale associated with short range correlations (SRCs) between nucleons. Therefore, the EMC effect is naturally connected with short range correlations between nucleons. On the other hand, the inter-nucleon separation is not much smaller than that of the nuclear size. This means that effects involving the entire nucleus cannot be disregarded. Such effects are known as mean-field effects in which each nucleon moves in the mean field provided by other nucleons. Thus, an explanation of the EMC effect should involve physics at all three length scales~\cite{Miller:2020eyc}.\\

The contents of this paper is as follows: In Sec. II, we will discuss the role of virtuality in motivating the use of quark degrees of freedom, as well as aid us in identifying the dominant interactions in the EMC effect. In Sec. III we will present a two-component model of the nucleon that will guide our intuition throughout this paper. Furthermore, this model of the nucleon will give us a relationship between virtuality and the nuclear potential. In Sec. IV we will present our construction of a new model of the EMC effect using holographic QCD. We will first summarize results from \cite{sufian_analysis_2017,deTeramond:2018ecg}, giving a framework in obtaining free nucleon Parton Distribution Functions (PDFs) from free nucleon form factors. Then, we will introduce the effects of a nuclear medium, allowing one to obtain the modified nucleon PDFs. In Sec. V we will present the model's expressions for the EMC ratios show the model's results for the EMC ratios for a variety of nuclei. Sec. VI will present an argument that will identify the dominant interaction in the EMC effect, mean field or SRC. Lastly, Sec. VII will provide a check of the change in the electric charge radius of a nucleon inside a nucleus. Our concluding remarks are given in Sec. VIII
 
 
\section{Virtuality - a small-distance scale  }


Bound nucleons of four momentum $p$ do not obey the standard Einstein relation: $p_\m p^\m=p^2=M^2$, and are thus said to be off their mass shell. By examining the intermediate nucleons in nucleon-nucleon scattering, we can gain insight into why bound nucleons do not obey Einstein's relation. In the Blankenbecler-Sugar \cite{PhysRev.142.1051} and Thompson reductions~\cite{Thompson:1970wt} of the Bethe-Salpeter equation~\cite{Salpeter:1951sz}, one nucleon emits a meson of zero energy and non-zero momentum, while the other nucleon absorbs the meson. Since the momenta of the nucleons have changed, but their energies have not, $p^2 \neq M^2$, meaning the intermediate nucleons are off their mass shell. In other reductions of the Bethe-Salpeter equation \cite{Gross:1969rv}, one nucleon is on the mass shell, and the other is not. This means that the nuclear wave function, treated relativistically, contains nucleons that are off their mass shell. Such nucleons must undergo interactions before they can be observed, and are thus denoted as virtual, with difference $p^2-M^2$ being proportional to the virtuality, $\cal V$~\cite{Miller:2019mae}. Experiments~\cite{Egiyan:2003vg,Egiyan:2005hs,Fomin:2011ng} using leptonic probes at large values of Bjorken $x$ interrogate the virtuality of the bound nucleons. Plateaus, kinematically corresponding to scattering by a pair of closely connected nucleons, have been observed \cite{Fomin:2017ydn} in this region. \\


For a nucleon to be so far off the mass shell, it needs to be interacting strongly with another nearby nucleon. To see that, consider a configuration of two bound nucleons initially at rest in the nucleus. This is a good approximation for roughly 80\% of the nuclear wave function. To acquire a large virtuality, one nucleon must exchange a boson or bosons with four-momentum, $\vec P$, comparable to that of the incident virtual photon. 
Such a bosonic system can only travel a short distance $\D r$ between the nucleons with 

\begin{equation} \label{small}
    \Delta r \sim \frac{1}{|\vec P|}
\end{equation}
 thus a highly virtual nucleon gets its virtuality from another nearby nucleon which must be closely separated. High virtuality is a short-distance phenomenon, and as such will help us determine whether an interaction is being affected by SRCs. \\





Additionally, high virtuality serves as an intermediate step between using nucleonic and quark degrees of freedom. To better understand this, consider a virtual nucleon as a superposition of physical states that are eigenfunctions of the QCD Hamiltonian.
Virtual states with nucleon quantum numbers can be expressed using the completeness of states of QCD,
\bea |N(\Cv)\ra=\sum_{n=1}^{n_{\rm max}} c_n |N_n\ra, \label{sup} \eea in which 
the states $|N_n\ra$ are resonances and also nucleon-multi-pion states.   Each of these states (with the total three-momentum of the state $ |N(\Cv)\ra)$ has a detailed underlying structure in terms of quarks and gluons.
In  exclusive reactions  with not very large momentum transfer  few states are excited and  one may use \eq{sup} to describe the physics.  However, for high energy inclusive reactions of experimental relevance one needs many states. Because of the large number of states entering in \eq{sup} it is most efficient to use quark degrees of freedom to understand DIS large values of $Q^2$. Thus, the free nucleon can be regarded as a superposition of various configurations or Fock states, each with a different quark-gluon structure. 
 
\section{Two-Component Model of the Nucleon}
As discussed in Sec. II, it is most efficient to use quark degrees of freedom due to the large number of states entering Eq. (\ref{sup}). 
Motivated by the the QCD  physics of color transparency \cite{Frankfurt:1985cv,Brodsky:1987xw,Ralston:1988rb,Jennings:1993hw}, we will treat the infinite number of quark-gluon configurations of the nucleon as two configurations: a large-sized, blob-like configuration  (BLC), consisting of complicated configurations of many quarks and gluons, and a small-sized, point-like configuration (PLC) consisting of 3 quarks. The BLC can be thought of as an object that is similar to a nucleon, and the PLC is meant to represent a three-quark system of small size that is responsible for the high-$x$ behavior of the distribution function; the smaller the number of  quarks, the more likely one can carry a large momentum fraction.  \\ 


When placed in a nucleus, the blob-like configuration feels the
usual  nuclear attraction and its energy decreases. The
point-like-configuration feels far less nuclear-attraction   by virtue of color screening \cite{Frankfurt:1994hf}, in which the
effects of gluons emitted by small-sized configurations are cancelled in low-momentum transfer processes. The nuclear
attraction increases the energy difference between the BLC and the PLC, therefore reducing the PLC probability~\cite{Frankfurt:1985cv}. Reducing the PLC probability in the nucleus reduces the quark momenta, in qualitative agreement with the EMC effect. Working out the consequences of the BLC-PLC model enables the connection between the EMC effect and virtuality to be clarified.\\

\subsection{The Free Nucleon}
The Hamiltonian  for a free nucleon  in the two-component model can be expressed schematically by the matrix 
\bea H_0=
\left[ \begin{array}{cc}
    E_B & V \\
    V & E_P \end{array} \right] \eea 
where $P$ the PLC and $B$ represents BLC.
We define the energy difference between the PLC and the BLC to be $\D = E_P - E_B$.
The hard-interaction potential, $V$, connects the two
components, causing the eigenstates of $H_0$ to be  $|N\ra$ and $|X\ra$  rather
than $|B\ra$ and $|P\ra$. 
 The normalized
eigenstates are given by 

\begin{equation} \label{nuc0}
    |N\ra = \frac{1}{\sqrt{1+\epsilon^2}}(|B\ra  +\epsilon|P\ra),  
\end{equation}

\begin{equation}
    |X\ra = \frac{1}{\sqrt{1+\epsilon_X^2}}(|B\ra + \epsilon_X|P\ra).
\end{equation}

\noindent where 

\begin{equation}
    \epsilon =\frac{-2V}{\Delta + \bar\Delta} \, , \,\,\,\,\,\,\,\ \e_X=\frac{2V}{- \Delta + \bar\Delta} \, , \,\,\,\,\,\,\,\
      \bar\Delta\equiv\sqrt{\Delta^2+4V^2}
\end{equation}

\noindent The notation $|X\ra$  is used to denote the  orthogonal excited  state. For later use, the probability of the PLC, $P_{PLC}$, for free nucleon is

\begin{equation} \label{prob_nomod}
    P_{PLC}= \frac{\epsilon^2}{1 + \epsilon^2}.
\end{equation}


\subsection{Medium effects}
 Now suppose the nucleon is bound to a nucleus. The nucleon feels an attractive nuclear potential, here represented by  $H_{1(n,p)}$, with 
 \bea H_{1(n,p)}= \left[ \begin{array}{cc}
U_{(n,p)} & 0 \\
0 & 0  \end{array} \right]
 ,\label{H1}\eea
to represent the idea that  only the large-sized component of the nucleon feels the influence of the nuclear attraction. Note that $U_{(n,p)}$ is dependent on $A$ and $Z$. The treatment of the nuclear interaction, $U_{(n,p)}$,  as a number is clearly a simplification because 
the interaction necessarily varies with the relevant kinematics.\\

The complete Hamiltonian $H=H_0+H_{1(n,p)}$ is:
\bea H= \left[ \begin{array}{cc}
E_B-|U_{(n,p)}| & V \\
V & E_P \end{array} \right],
 \eea
in which the attractive nature of the nuclear  binding potential is
emphasized. Then interactions with the nucleus increase the energy difference between
the bare BLC and  PLC states and thereby decrease the PLC
probability.  \\


The medium-modified nucleon and its excited state,
$|\tilde{N}\ra$ and $|\tilde{X}\ra$, are now 
\begin{equation} 
    |\tilde{N}\ra = \frac{1}{\sqrt{1+\tilde{\epsilon}^2}}(|B\ra  +\tilde{\epsilon}|P\ra),  
\end{equation}

\begin{equation}
    |\tilde{X}\ra = \frac{1}{\sqrt{1+\tilde{\epsilon}_X^2}}(|B\ra + \tilde{\epsilon}_X|P\ra).
\end{equation}

The expression for $\tilde{\epsilon}$ can be obtained by making the replacement: $\Delta \rightarrow \Delta + |U_{(n,p)}|$

\begin{equation} \label{e_mod_no_expansion}
       \tilde{\epsilon}={- 2V\over \D + |U_{(n,p)}| + \sqrt{(\D + |U_{(n,p)}|)^2 + 4V^2}} 
\end{equation}

Since $V$ is associated with the strong force, and $U_{(n,p)}$ to the nuclear force, we can expand Eq. (\ref{e_mod_no_expansion}) to first order in $\frac{|U_{(n,p)}|}{\D}$,
\be
\tilde{\e}_{(n,p)} \approx \e(1 - {|U_{(n,p)}|\over \bar\D}) \label{epsilon_mod}
\ee

\noindent The probability of the modified PLC for the nucleon, $\tilde{P}_{PLC}(n,p)$, is now

\begin{equation} \label{pdef1}
    \tilde{P}_{PLC}(n,p)= {\tilde{\e}_{(n,p)}^2\over 1+\tilde{\e}_{(n,p)}^2}
\end{equation}


\noindent Replacing $\tilde{\epsilon}_{(n,p)}$ with Eq. (\ref{epsilon_mod}), expanding to first order in $\frac{|U_{(n,p)}|}{\bar{\D}}$, and solving for $\tilde{r}_{(n,p)}$ we get

\begin{equation} \label{prob_mod}
 \tilde{P}_{PLC}(n,p)={\e^2\over 1 + \e^2} - \frac{2 |U_{(n,p)}|  \e^2}{\bar{\D} (1 + \e^2)^2} = P_{PLC} \left(  1 - \frac{2|U_{(n,p)}|}{\bar{\D} (1 + \epsilon^2)  } \right)
\end{equation}.

\subsection{Connecting the Nuclear Potential to Virtuality}
The next step is to relate $  U_{(n,p)}$  to the virtuality which is done in Ref. \cite{CiofidegliAtti:2007ork}. 
 Suppose a photon interacts with a virtual nucleon of four-momentum $ \bfP_{\rm miss}$.
 The three-momentum
 $\bfP_{\rm miss}$
opposes the $A-1$ recoil momentum $\bfp\equiv \bfP_{\rm miss}=-\bfP_{A-1}$. The  mass of the on-shell recoiling nucleus is given by
  $M_{A-1}^*=M_A-M+E,$ where $E>0$ represents the excitation energy of
  the spectator $A-1$ nucleus, and $M$ is the mass of the nucleon.
  \begin{eqnarray}
 &M^2 \Cv=
 \Pm^2-M^2\\&
 =(M_A-\sqrt{(M_{A-1}^*)^2 +\bfp ^2}\,\,)^2
- \bfp^2-M^2
 \end{eqnarray}
which reduces  in the non-relativistic limit to
  \begin{eqnarray}
M^2 \Cv_{(n,p)}&\approx& -2M\left({\bfp^2\over 2M_r}+E_{(n,p)}\right),
\label{virt}
\end{eqnarray}

where the reduced mass  $M_r= M(A-1)/A$. The virtuality,  $\Cv(n,p)$,  is less than 0, and its magnitude increases with both the $A-1$ excitation energy and the  initial momentum of the struck nucleon.

Refs.~\cite{Frankfurt:1985cv,CiofidegliAtti:2007ork} obtained a relation between
the potential $U_{(n,p)}$ and the virtuality $\Cv_{(n,p)}$ by using the extension of the Schroedinger equation to an
operator form:
\bea 
{\bfp^2\over 2M_r}+U_{(n,p)}=-E_{(n,p)},
\eea

so that $ {\bfp^2\over 2M_r}+E_{(n,p)}=-U_{(n,p)}=|U_{(n,p)}|$, we get

\bea \label{cv}
\Cv_{(n,p)} = {-2|U_{(n,p)}|\over M},
\eea

\noindent thus directly connecting the nuclear potential and the virtuality.

\section{Medium Modification in LFHQCD}
We will now introduce a new model of the EMC effect using holographic QCD nucleon form factors as a starting point \cite{sufian_analysis_2017}. The SU(6) spin-flavor symmetric quark model is used in calculating the effective charges of positive and negative helicity protons and neutrons. These effective charges are used to obtain expressions for the nucleon form factors. In order to extend this model to nuclei, the key idea is that the nuclear medium will affect the probabilities of finding a spin up or down quark $q$ in a proton or neutron, breaking SU(6) symmetry; thus, the effective charges are ultimately modified as well. We will use intuition from the two-component model of the nucleon to guide us in parameterizing the new modified effective charges. Furthermore, connecting to arguments presented in Sec. II, the formalism of LFHQCD presented in Ref. \cite{sufian_analysis_2017} allows us to write valence free and modified PDFs in terms of quark degrees of freedom.  

\subsection{Free Nucleon PDFs from Holographic QCD}

To summarize results in \cite{sufian_analysis_2017}, in LFHQCD the electromagnetic form factor for an arbitrary twist-$\tau$ hadron is

\begin{equation} \label{hff}
    F_{\tau}(Q^2) = \int \frac{dz}{z^3} V(Q^2,z)\Phi_{\tau}^2
\end{equation}

\begin{equation} \label{hlfwf}
     \Phi_{\tau}(z) = \sqrt{\frac{2}{\Gamma(\tau - 1)}} \kappa^{\tau - 1}z^{\tau} e^{\kappa^2 z^2/2}
\end{equation}

\noindent where $V(Q^2,z)$ is the bulk to boundary propogator, and $\Phi_{\tau} (z)$ is the twist-$\tau$ hadronic wave function. The spin-nonflip elastic Dirac form factor for a nucleon $N$, $F_1^N$, is given by

\begin{equation} \label{F1N}
F_1^N(Q^2)  = \sum_{\pm} g_{\pm}^N \int \frac{dz}{z^4} V(Q^2,z) \Psi_{\pm}^2 (z)
\end{equation}

\noindent with

\begin{equation}\label{gpm}
\begin{split}
    g_{+}^N = P_{N \uparrow}^u e_u + P_{N \uparrow}^d e_d 
    \\
    g_{-}^N = P_{N \downarrow}^u e_u + P_{N \downarrow}^d e_d,
\end{split}
\end{equation}

\noindent where $g_{\pm}^N$ are the effective charges for a positive ($+$) or negative ($-$) chirality nucleon $N$, $P_{N (\uparrow,\downarrow)}^{q}$ is the probability to find a spin up or down quark $q$ in a nucleon $N$, $e_q$ is the charge of a quark $q$ in units of positron charge $e$, and $\Psi_{\pm} (z)$ are the wave functions corresponding to a positive ($+$) or negative ($-$) chirality nucleon. Notably, $\Psi_{\pm} (z)$ have the following dependencies,

\begin{equation}\label{nlfwf}
\begin{split}
    \Psi_+(z) \sim z^{\tau + \frac{1}{2}} e^{\kappa^2 z^2/2} \,\,\,\,\,
    \\
    \Psi_-(z) \sim z^{\tau + 1 + \frac{1}{2}} e^{\kappa^2 z^2/2}.
\end{split}
\end{equation}

The SU(6) symmetry approximation is used order to obtain $P_{N (\uparrow,\downarrow)}^{q}$. Using this symmetry approximation, the effective charges become

\begin{equation}
    g_+^p = 1, \,\,\ g_-^p = 0, \,\,\ g_+^n = -\frac{1}{3}, \,\,\ g_-^n = \frac{1}{3}.
\end{equation}

Ref. \cite{sufian_analysis_2017} next introduces a free parameter, $r$, that multiplies the neutron effective charges, $g_+^n$ and $g_-^n$, in order to properly match $F_1^n$ to existing experimental data. In this paper, we will use $r = 1.5$, as done in Ref. \cite{deTeramond:2018ecg} and further motivated by an argument using wave function normalization \cite{privateconvo}. The effective charges now become 

\begin{equation} \label{eff_charge_nomod}
    g_+^p = 1, \,\,\ g_-^p = 0, \,\,\ g_+^n = -\frac{1}{2}, \,\,\ g_-^n = \frac{1}{2}.
\end{equation}

In order to obtain expressions for the nucleon form factors, a simplified model is introduced which only uses the leading twist-3 term in the nucleon wave function. This leads to the following results for $F_1^N$,

\begin{equation} \label{f1p_nomod}
    F_1^p(Q^2)  = F_{\tau=3}(Q^2) 
\end{equation}

\begin{equation} \label{f1n_nomod}
    F_1^n(Q^2)  = -\frac{1}{2}F_{\tau=3}(Q^2)  + \frac{1}{2} F_{\tau = 4}(Q^2).
\end{equation}

One can obtain the up ($u$) and down ($d$) valence PDFs of the free proton and neutron by using a flavor decomposition of nucleon form factors \cite{cates_flavor_2011}, and by writing the form factors for quark flavor $q$, $F^q$, in terms of the valence GPD $H^q_v (x,t)$,

\begin{equation} \label{flavordecomp}
    F_1^N  = \frac{2}{3} (F_1^u)^N - \frac{1}{3}(F_1^d)^N,
\end{equation}

\begin{equation} \label{flavordecomp_isospin}
    F_1^N = \frac{2}{3} (F_1^u)^N - \frac{1}{3}(F_1^d)^N \xrightarrow[\text{Symmetry}]{\text{Isospin}} (F_1^u)^p = 2F_1^p + F_1^n, \,\,\ (F_1^d)^p = F_1^p + 2F_1^n,
\end{equation}

\begin{equation} \label{flavorff}
    (F^q_1)^N = \int_0^1 dx \, H^q_v (x,t) = \int_0^1 dx \, q_v^N(x) e^{t f(x)},
\end{equation}

\noindent where $(F_1^q)^N$ is the $F_1$ flavor form factor for quark $q$ in nucleon $N$, $q_v^N(x)$ is the valence PDF for quark flavor $q$ in nucleon $N$, and $f(x)$ is the profile function \cite{deTeramond:2018ecg}. 

Furthermore, Ref. \cite{de_teramond_gaugegravity_2010} recast Eq. (\ref{hff}) in terms of an Euler Beta Function and determined what the PDF is for $(F^q_1)^N = F_\tau$. The corresponding PDF for $F_\tau$, referred to as $q_\tau(x)$, is normalized to unity and is given by,

\bea  &
q_\tau(x) =\frac{\Gamma \left(\tau -\frac{1}{2}\right)}{\sqrt{\pi } \Gamma (\tau -1)}\big(1- w(x)\big)^{\tau-2}\, w(x)^{- \half}\, w'(x), \label{qt}
  \eea

\noindent with \bea w(x) = x^{1-x} e^{-a (1-x)^2} \eea

\noindent where the flavor-independent parameter $a = 0.531 \pm 0.037$. Using Eqs. (\ref{f1p_nomod}, \ref{f1n_nomod}, \ref{flavordecomp_isospin}, \ref{flavorff}) one can obtain the valence $u$ and $d$ proton quark distributions at the matching scale between LFHQCD and pQCD, $\m_0=1.06\pm 0.15 $ GeV \cite{deTeramond:2018ecg}.
\bea &
u_v^p(x) ={3\over 2}q_3(x) +{1\over 2} q_4(x)\nonumber\\
&d_v^p(x)=q_4(x) \label{qdist}
,\eea
One can obtain the neutron valence PDFs through isospin symmetry. 



From now on, we will drop the subscript $v$ and it is implied that all PDFs presented in this paper are valence. Also, notice that the above PDFs are expressed as a superposition of twist-$\tau$ PDFs, our quark degrees of freedom. The square of the proton and neutron wave functions are characterized by,

\begin{equation} \label{wfp_nomod}
    \Psi_p^2 \sim u^{p} + d^{p} = \frac{3}{2}q_3(x) + \frac{3}{2}q_4(x)
\end{equation}

\begin{equation}\label{wfn_nomod}
    \Psi_n^2 \sim u^{n} + d^{n} = \frac{3}{2}q_3(x) + \frac{3}{2}q_4(x)
\end{equation}

\begin{figure}[H]
\hspace*{2.5cm}\includegraphics[width=5in]{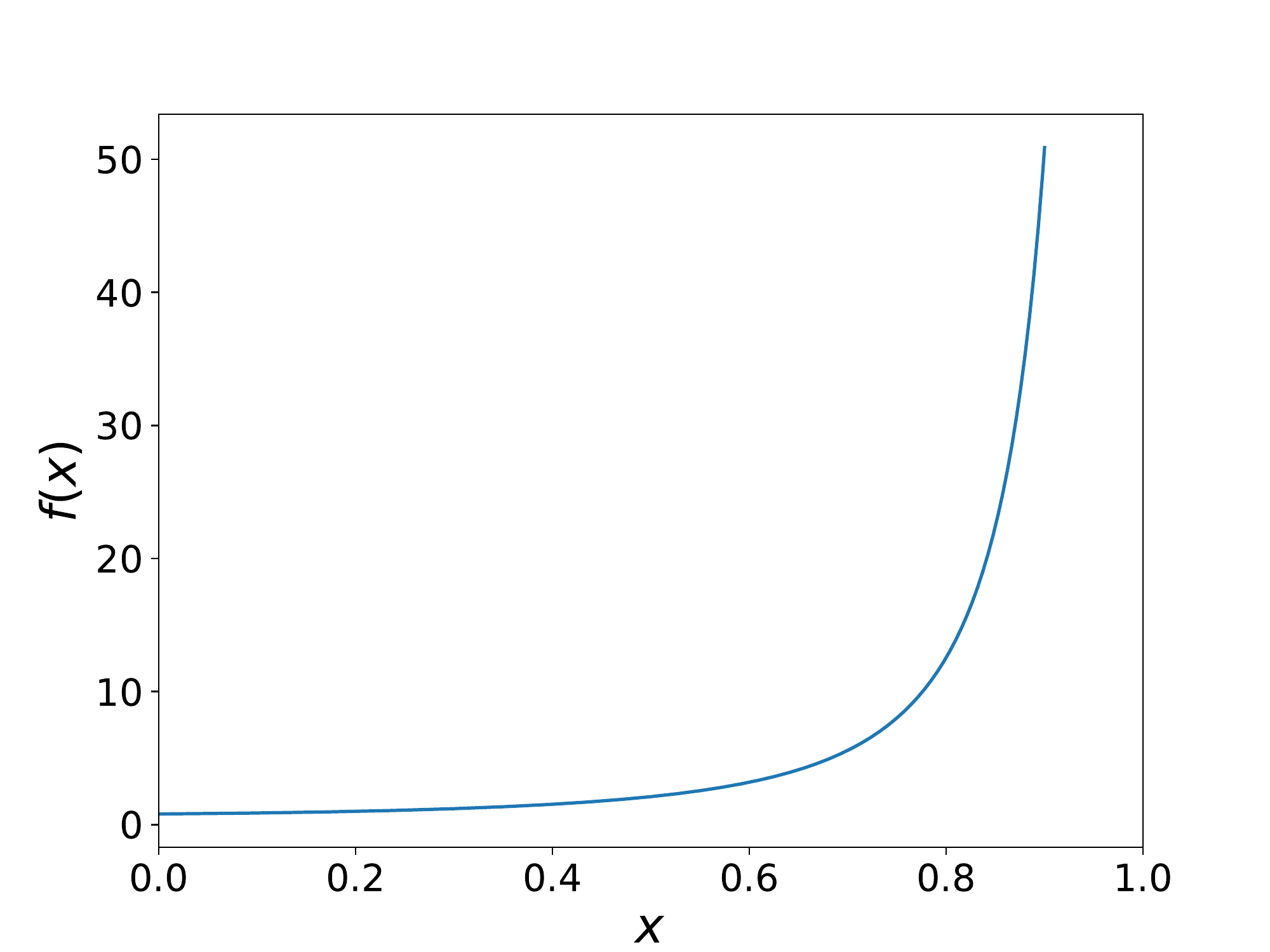}
\caption{A plot of $f(x) = q_3/q_4$, as a function of $x$. Notice that $f(x)$ increases with increasing $x$, displaying the PLC dominance at high-$x$.} 
\label{fig:fxfig}
\end{figure}

Tying back to the two-component model, the elastic form factors in the LFHQCD model fall asymptotically as $1/Q^{2\t}$, and the slope of form factors as $Q^2=0$ is proportional to $\t$. These features  mean  that an increase in the value of $\t$ corresponds to an increase in effective size. Furthermore, Ref. \cite{deTeramond:2018ecg} notes that $\tau$ refers to the number of constituents in a given Fock component of the hadron. Therefore, the function $q_3$ is naturally associated with the a three quark PLC system and $q_4$ with the BLC. This association is also consistent with the discussion regarding the PLC dominating at high-$x$ as can be seen by analyzing $f(x) = q_3/q_4$ (Fig. \ref{fig:fxfig}).

Lastly, normalizing Eqs. (\ref{wfp_nomod}, \ref{wfn_nomod}), the probability of the PLC, $P_{PLC}$, for both nucleons is equal to 1/2. Using Eq. (\ref{prob_nomod}), we find that $\epsilon = -1$.

\subsection{Modified Nucleon PDFs From Holographic QCD}
In order to introduce the effects of a nuclear medium we must identify terms in which the medium would modify: these terms are the probabilities that go into calculating the effective charges, $P_{N (\uparrow \downarrow)}^q$, and the nucleon wave functions, $\Psi_{\pm}(z)$. To obtain the nucleon wave functions, one must solve the effective single-variable light-front Schr\"{o}dinger Equation (SE) \cite{de_teramond_gaugegravity_2010}. The SE involves an effective potential that encompasses the confining interaction terms in the QCD Lagrangian, i.e. the potential due to the strong force. On the other hand, the nuclear medium can be thought of as the potential due to the nuclear force. As a result, the modification to the effective potential due to a nuclear medium is small; the same is true for $P_{N (\uparrow \downarrow)}^q$ by similar reasoning. In this study, we will only consider the consequences of nuclear mediums breaking SU(6) symmetry - i.e. modifying $P_{N (\uparrow \downarrow)}^q$. 

Motivated by np dominance in SRCs, we expect the nuclear potential to depend on whether one introduces a proton or neutron into the nucleus. For example, we expect the proton to feel a stronger attraction to a nucleus if there is an abundance of neutrons and vice versa. We apply medium effects by introducing two free parameters (which both depend on mass and atomic numbers $A$ and $Z$ respectively), $\delta r_p(A,Z)$ and $\delta r_n(A,Z)$. With these two phenomenological parameters, we parameterize the effective charges in Eq. (\ref{eff_charge_nomod}) as

\begin{equation} \label{eff_charge_mod}
    \tilde{g}_+^p = 1 - \delta r_p(A,Z), \,\,\ \tilde{g}_-^p = \delta r_p(A,Z), \,\,\ \tilde{g}_+^n = -\frac{1}{2} - \delta r_n(A,Z), \,\,\ \tilde{g}_-^n = \frac{1}{2} + \delta r_n(A,Z).
\end{equation}

\noindent The $A$ and $Z$ dependencies in $\delta r_p$ and $\delta r_p$ are dropped from now on and are implied. The signs in front of $\delta r_p$ and $\delta r_n$ are motivated by the suppression of the PLC from the two-component model. We will soon see that the above parameterization leads to a suppression of the PLC, $q_3(x)$. Notice that if there is no nuclear medium, $\delta r_p$ = $\delta r_n$ = $0$, and the free nucleon effective charges are regained.

With Eq. (\ref{eff_charge_mod}), Eqs. (\ref{f1p_nomod}, \ref{f1n_nomod}) now become

\begin{equation} \label{f1p_mod}
    \tilde{F}_1^p(Q^2)  = (1 - \delta r_p)F_{\tau=3}(Q^2)  + \delta r_p F_{\tau=4}(Q^2),
\end{equation}

\begin{equation} \label{f1n_mod}
     \tilde{F}_1^n(Q^2)  = -\left(\frac{1}{2} + \delta r_n\right)F_{\tau=3}(Q^2)  + \left(\frac{1}{2} + \delta r_n\right) F_{\tau = 4}(Q^2).
\end{equation}

Care must be taken in obtaining the modified nucleon $u$ and $d$ valence PDFs. For $N \neq Z$, one cannot use Eq. (\ref{flavordecomp_isospin}) as the nuclear medium modifies protons and neutrons differently. However, one can use Eq. (\ref{flavordecomp_isospin}) for $N = Z$. We can thus obtain the expressions of modified nucleon valence PDFs for $N = Z$ and use their forms to intuit expressions for what the valence PDFs should be for arbitrary $N$ and $Z$. This process gives us the following medium modified proton valence PDFs,

\begin{equation} \label{mod_p_u}
    \tilde{u}^p = \left(\frac{3}{2} - 3\delta r_p\right)q_3(x) + \left(\frac{1}{2} + 3\delta r_p\right)q_4(x)
\end{equation}

\begin{equation} \label{mod_p_d}
    \tilde{d}^p = \left( - 3\delta  r_p\right)q_3(x) + \left(1 + 3\delta r_p\right)q_4(x)
\end{equation}

\noindent and the following modified neutron valence PDFs.

\begin{equation} \label{mod_n_u}
    \tilde{u}^n = \left( - 3\delta  r_n\right)q_3(x) +\left(1 + 3\delta r_n\right)q_4(x)
\end{equation}

\begin{equation} \label{mod_n_d}
    \tilde{d}^n = \left(\frac{3}{2} - 3\delta r_n\right)q_3(x) + \left(\frac{1}{2} + 3\delta  r_n\right)q_4(x)
\end{equation}

One can check that the above expressions for the modified PDFs, using Eqs. (\ref{flavordecomp}, \ref{flavorff}), give Eqs. (\ref{f1p_mod}, \ref{f1n_mod}). Again, notice that the above modified PDFs are expressed as a superposition quark degrees of freedom, $q_\tau$. Further, notice that we have a suppression of the PLC contribution to the above modified nucleon valence PDFs. 

It is important to note that Eqs. (\ref{mod_p_u}, \ref{mod_p_d}, \ref{mod_n_u}, \ref{mod_n_d}) are not quantities constrained by data, the quantites that experimental DIS data constrains are the nuclear PDFs, $f^A$. Motivated by Ref. \cite{kovarik_ncteq15_2016}, we will define the nuclear PDFs as

\begin{equation}
    f^{A} = Z \tilde{f}^p + N \tilde{f}^n,
\end{equation}

\noindent Where $f$ denotes the quark flavor, $Z$ is the atomic number, $N$ is the number of neutrons, and $\tilde{f}^p$ ($\tilde{f}^n$) is the modified proton (neutron) PDF in nucleus $A$. 

Lastly, the square of the modified proton and neutron wave functions are characterized by,

\begin{equation} \label{wfp_mod}
    \tilde{\Psi}_p^2 \sim \tilde{u}^p + \tilde{d}^p = \left(\frac{3}{2} - 6\delta r_p \right) q_3(x) + \left(\frac{3}{2} + 6 \delta r_p \right) q_4(x)
\end{equation}

\begin{equation} \label{wfn_mod}
    \tilde{\Psi}_n^2 \sim \tilde{u}^n + \tilde{d}^n = \left(\frac{3}{2} - 6\delta r_n \right) q_3(x) + \left(\frac{3}{2} + 6 \delta r_n \right) q_4(x)
\end{equation}

\noindent where even though we ignored the effects of the nuclear medium on $\Psi^{\pm}$, the wave functions still get modified due to modifications in the effective charges. Normalizing Eqs. (\ref{wfp_mod}, \ref{wfn_mod}), we find that the modified PLC probability is

\begin{equation}
    \tilde{P}_{PLC} = \frac{1}{2} - 2\delta r_p.
\end{equation}

Using Eq. (\ref{prob_mod}) and noting that $P_{PLC} = 1/2$, which leads to $\epsilon = 1$, we get a relationship between $\delta r_{(n,p)}$ and the nuclear potential,

\begin{equation}
    \tilde{P}_{PLC} = \frac{1}{2} - 2\delta r_p  = \frac{1}{2} \left(  1 - \frac{|U_{(n,p)}|}{\bar{\D}} \right),
\end{equation}

\begin{equation} \label{deltar}
    \delta r_{(n,p)} = \frac{1}{4} \frac{|U_{(n,p)}|}{\bar{\D}}.
\end{equation}

Lastly, using Eq. (\ref{cv}) we can obtain a relationship between $\delta r_{(n,p)}$ and virtuality

\begin{equation} \label{deltar_relations}
    \delta r_{(n,p)} = \frac{1}{4} \frac{|U_{(n,p)}|}{ \bar{\Delta}}=- \frac{M}{8} \frac{ \Cv_{(n,p)}}{\bar{\Delta}}.
\end{equation}

\section{EMC Ratios}
Rewriting the modified proton valence PDFs in Eqs. (\ref{mod_p_u}, \ref{mod_p_d}), 

\begin{equation}
    \tilde u^p = u^p +3 \delta r_p (q_4 - q_3)
\end{equation}

\begin{equation}
    \tilde d^p = d^p +3 \delta r_p (q_4 - q_3),
\end{equation}

\noindent and vice versa for the modified neutron PDFs. The modified DIS structure function for the proton is,

\begin{equation}
    \frac{\tilde F_2^p}{x} = \frac{4}{9} \tilde u^p + \frac{1}{9} \tilde d^p 
\end{equation}

\begin{equation}
       \frac{\tilde F_2^p}{x} = \frac{F_2^p}{x} + \frac{5}{3} \delta r_p (q_4 - q_3) 
\end{equation}


The modified DIS structure function for the neutron is obtained the same way:

\begin{equation}
       \frac{\tilde F_2^n}{x} = \frac{F_2^n}{x} + \frac{5}{3} \delta r_n (q_4 - q_3).
\end{equation}


Thus, the $F_2$ DIS structure function for a nucleus of mass number A, with Z protons and N neutrons, is

\begin{equation}
    F_2^A = Z \tilde F_2^p + N \tilde F_2^n = \frac{4}{9} x u^A +  \frac{1}{9} x d^A,
\end{equation}

\begin{equation}
     F_2^A = Z F_2^p + N F_2^n + \frac{5x}{3}(q_4 - q_3 )(Z \delta r_p + N \delta r_n).
\end{equation}


The EMC ratio for deuterium is thus,

\begin{equation}
    \frac{F_2^d}{F_2^p + F_2^n} = 1 + 4 \, \delta r(^2H)\left( \frac{1 - f(x)}{1 + f(x)} \right),
\end{equation}

\noindent where $\delta r(^2H)$ is the value of $\delta r_p$ and $\delta r_n$ for deuterium. Lastly, the EMC ratio, relative to deuterium, for a nucleus of mass and atomic numbers A and Z is

\begin{equation}
     \frac{2}{A} \frac{F_2^A}{F_2^{d}} = \frac{2}{A} \frac{Z F_2^p + N F_2^n + \frac{5x}{3}(Z \delta r_p + N \delta r_n)(q_4 - q_3 )}{F_2^p + F_2^n + \frac{10x}{3}\delta r(^2H)(q_4 - q_3 )}.
\end{equation}

\subsection{Fitting}
We determined $\delta r_p$ and $\delta r_p$ for a variety of nuclei by performing a $\chi^2$ minimization procedure. 

\begin{equation} \label{chi}
    \chi^2 = \sum_{exp} \sum_{i} \frac{ ( f(x_{i}) - y_{i} * \eta_{exp})^2}{\sigma_{i^2}}
\end{equation}

We have a sum over experiments because we have data from different experiments for the same EMC ratio measurement. It is implied that the sum over $i$ is for the given experiment that is being summed over. $f(x_{i})$ is the model's prediction for the EMC effect evaluated at Bjorken $x_i$, $y_i$ is the experimental EMC effect data measured at Bjorken $x_i$, $\sigma_i$ is the uncertainty in the measurement of the EMC effect at Bjorken $x_i$, and $\eta_{exp}$ is the normalization factor that multiplies every data point in a given experiment. The inclusion of $\eta_{exp}$ is to further optimize the fitting of our model to experimental data. Most experiments include a normalization uncertainty, meaning that all measured data points can be multiplied by a constant that is within the normalization uncertainty. In addition to $\delta r_{p}$ and $\delta r_{p}$, $\eta_{exp}$ is also used as a fitting parameter with its fitting bounds being the normalization uncertainty for the given experiment being summed over. 

Furthermore, uncertainties in $\delta r_p$ and $\delta r_n$ due to uncertainties outside of fitting, e.g. uncertainties in experimental data, were obtained through Monte Carlo error propagation. The uncertainties due to fitting were accounted for by adding them to the Monte Carlo uncertainties in quadrature, and then taking the square root.

The fitting procedure done in this paper goes as follows: we first fit the deuterium EMC ratio data in order to obtain $\delta r(^2H)$. Using this value of $\delta r(^2H)$, we then fit the rest of the EMC ratio data we had. For the case with $^3He/^2H$ and $^3He/^3H$, we performed a simultaneous fitting in order to obtain $\delta r_p$ and $\delta r_n$ values for $^3He$ and $^3H$.

\subsection{Results}
For the figures in this section, the published data from SLAC was obtained from Ref. \cite{Gomez:1993ri}, JLab from Refs. \cite{jlabdata1,jlabdata2}, CLAS from Ref. \cite{Schmookler:2019nvf}, MARATHON from Ref. \cite{abrams_measurement_2022}, and BONuS from Ref. \cite{griffioen_measurement_2015}. Furthermore, we removed all isoscalar corrections in all experimental data used in this paper.

\begin{figure}[H]
\hspace{0.7in}\includegraphics[width=5in]{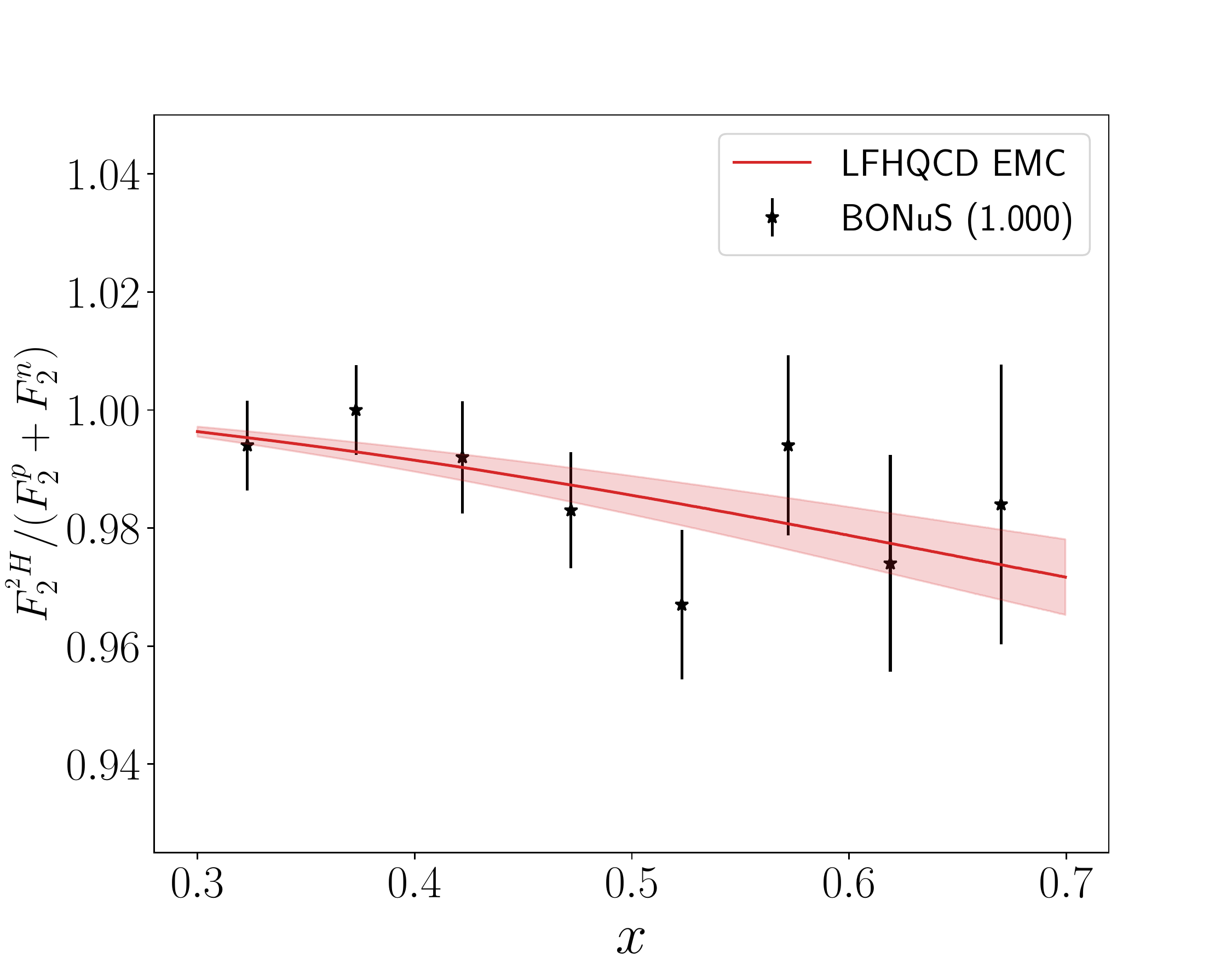}
\caption{Deuterium EMC ratio comparisons between the LFHQCD model (red line) and published experimental data (removed isoscalar corrections) obtained from BONuS data (filled stars). The red bands display 1$\sigma$ uncertainties for the LFHQCD EMC model. The number in parenthesis next to the experiment name in the legend is the normalization factor that multiplies all the data points, $\eta_{exp}$ in Eq. (\ref{chi}).} 
\label{fig:deutplot}
\end{figure}

\begin{figure}[H]
\includegraphics[width=6.5in]{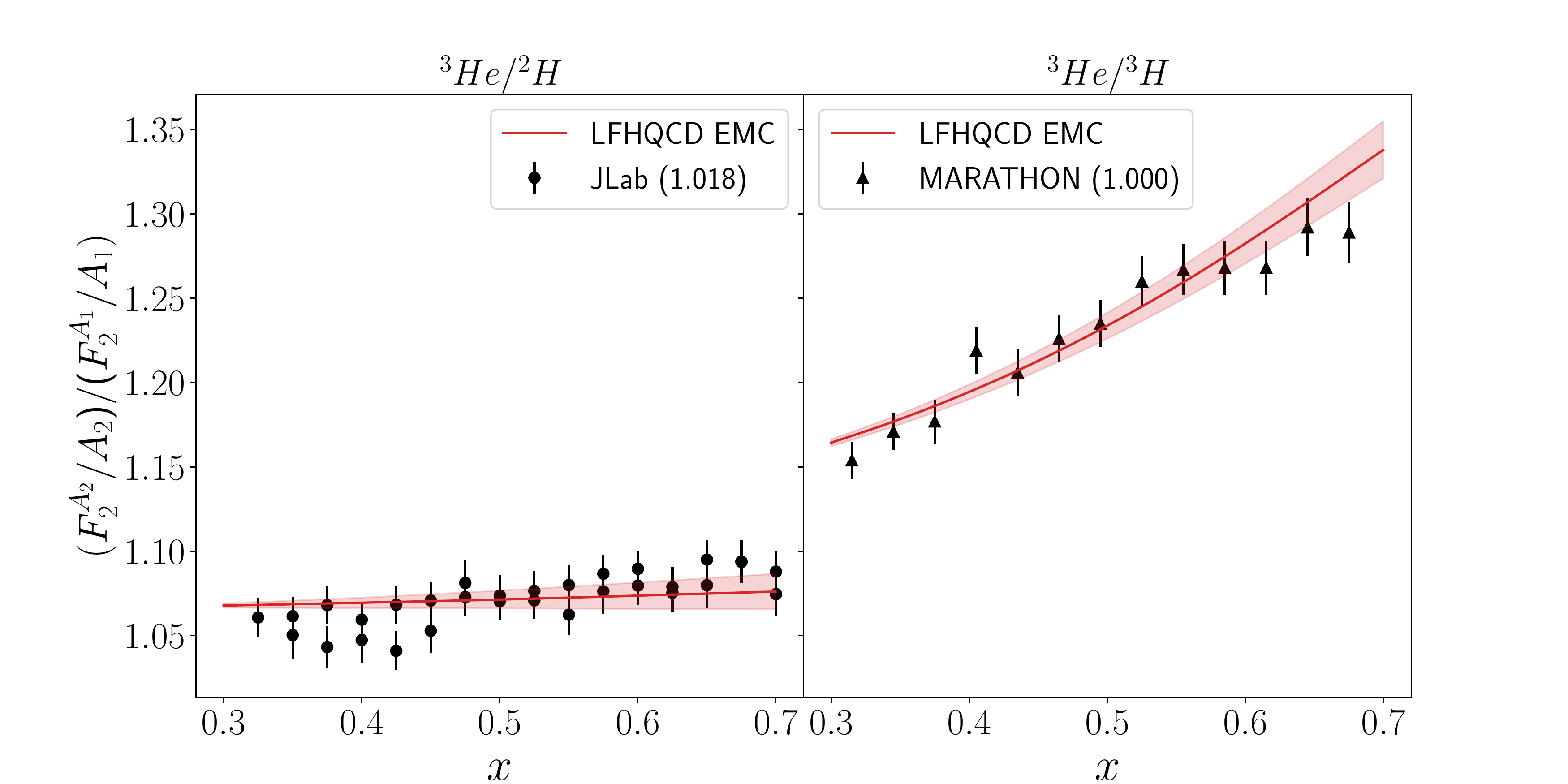}
\caption{EMC ratio comparisons between the LFHQCD model (red line) and published experimental data (removed isoscalar corrections) obtained from JLab (solid points) and MARATHON (solid triangles). The red bands display 1$\sigma$ uncertainties for the LFHQCD EMC model. The number in parenthesis next to the experiment name in the legend is the normalization factor that multiplies all the data points, $\eta_{exp}$ in Eq. (\ref{chi}).} 
\label{fig:plt1}
\end{figure}

\begin{figure}[H]
\hspace{-2.5cm}\includegraphics[width=8.75in]{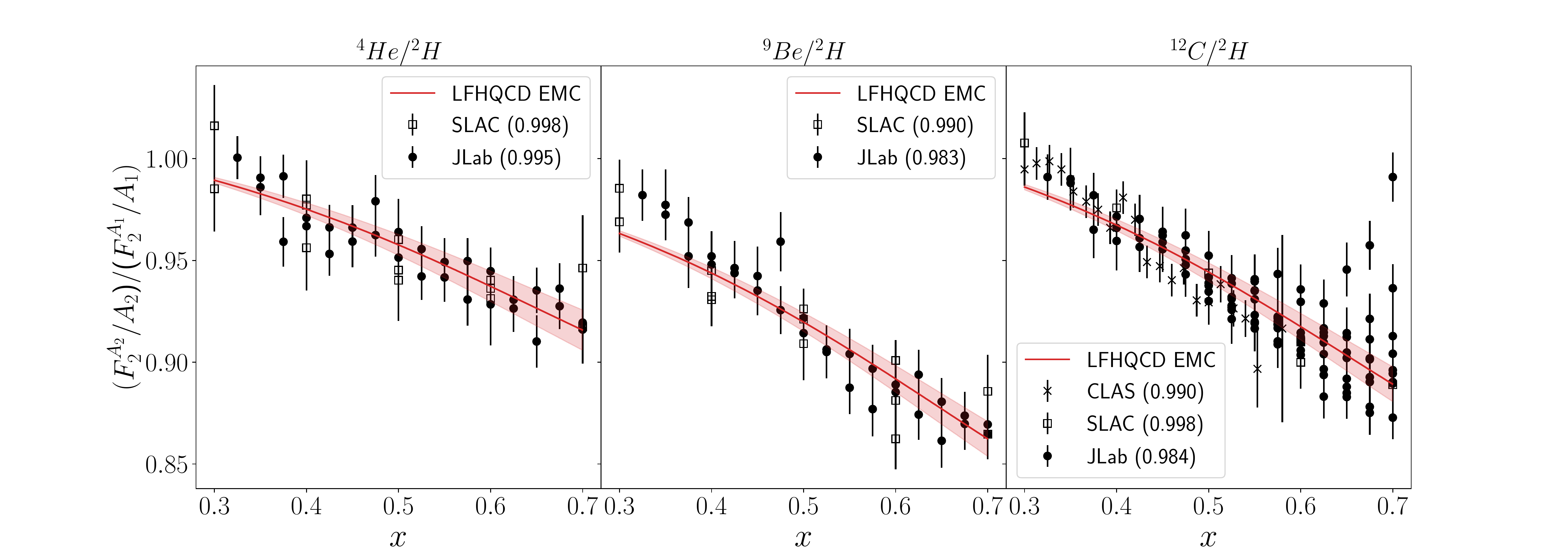}
\caption{EMC ratio comparisons between the LFHQCD model (red line) and published experimental data (removed isoscalar corrections) obtained from SLAC (open boxes), JLab (solid points), and CLAS (crosses). The red bands display 1$\sigma$ uncertainties for the LFHQCD EMC model. The number in parenthesis next to the experiment name in the legend is the normalization factor that multiplies all the data points, $\eta_{exp}$ in Eq. (\ref{chi}).} 
\label{fig:plt2}
\end{figure}

\begin{figure}[H]
\hspace{-2.5cm}\includegraphics[width=8.75in]{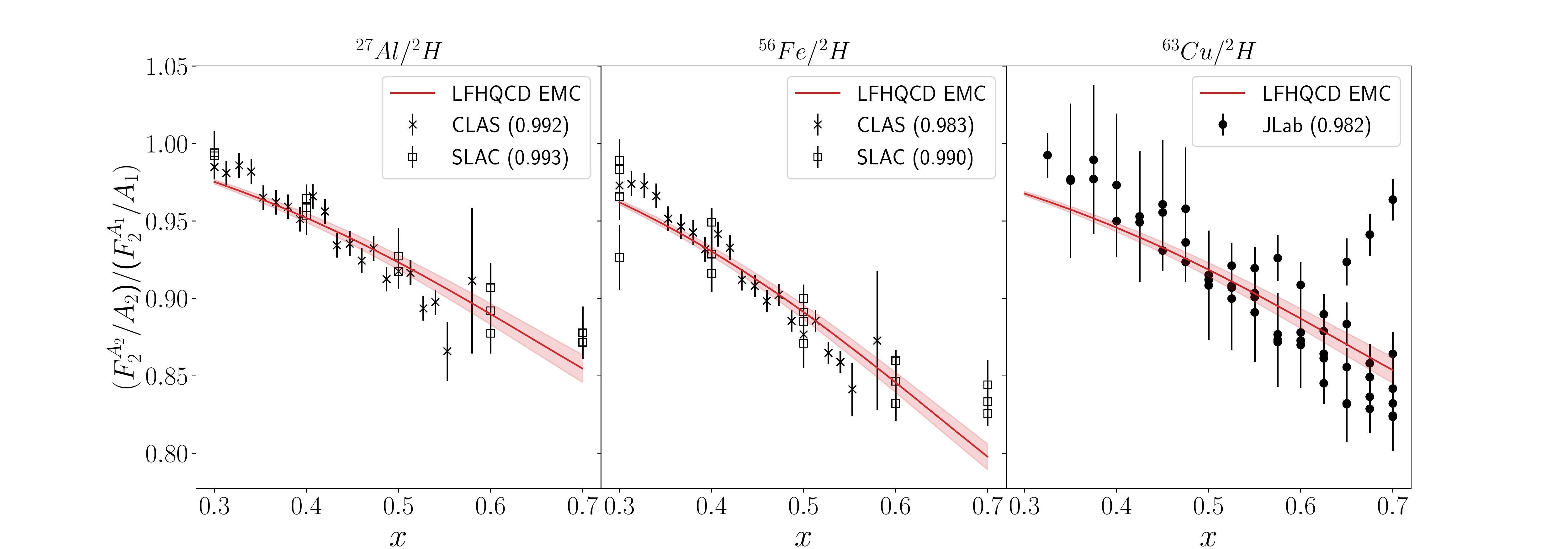}
\caption{EMC ratio comparisons between the LFHQCD model (red line) and published experimental data (removed isoscalar corrections) obtained from SLAC (open boxes), JLab (solid points), and CLAS (crosses). The red bands display 1$\sigma$ uncertainties for the LFHQCD EMC model. The number in parenthesis next to the experiment name in the legend is the normalization factor that multiplies all the data points, $\eta_{exp}$ in Eq. (\ref{chi}).} 
\label{fig:plt3}
\end{figure}

\begin{figure}[H]
\includegraphics[width=6.5in]{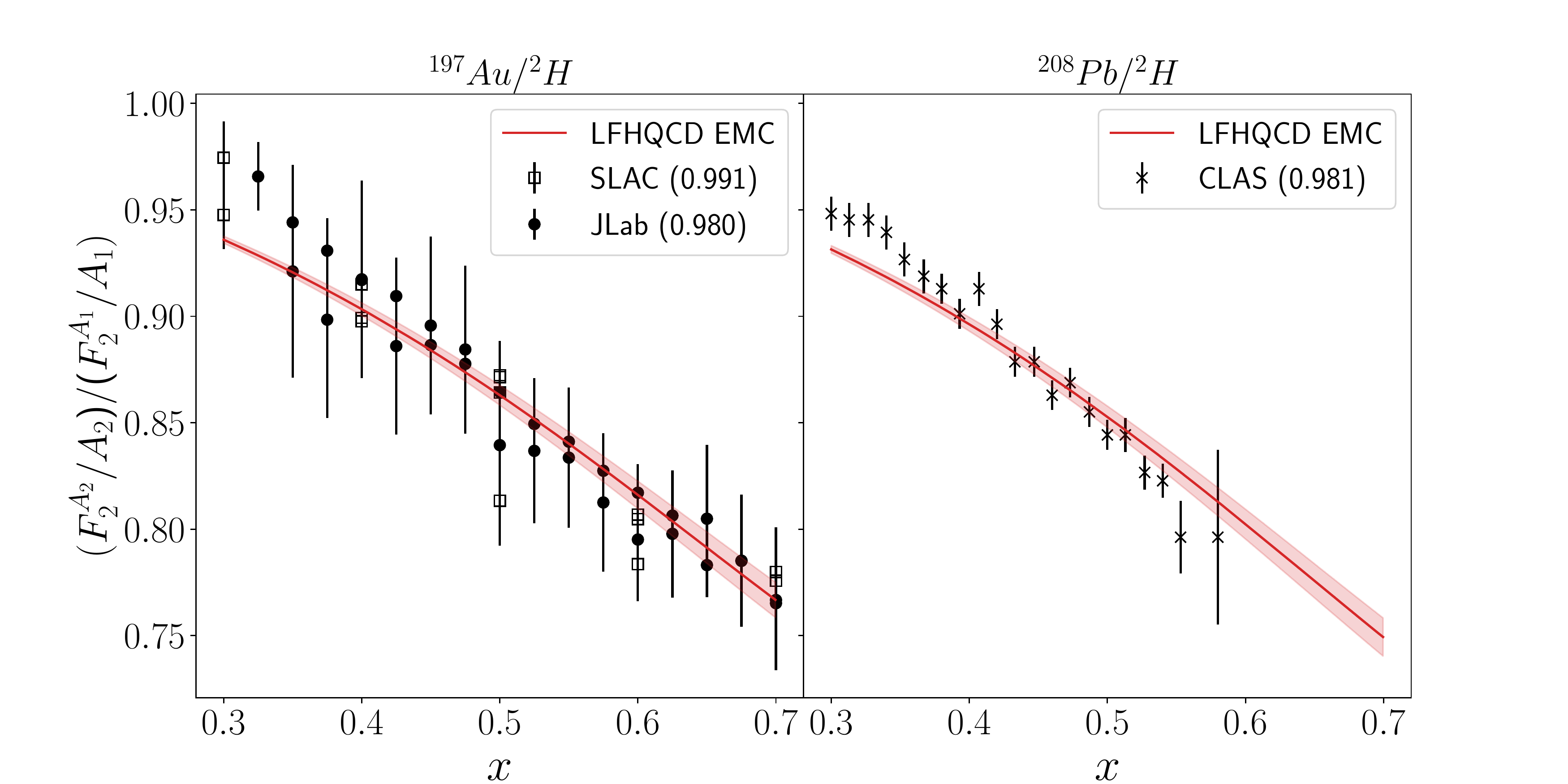}
\caption{EMC ratio comparisons between the LFHQCD model (red line) and published experimental data (removed isoscalar corrections) obtained from SLAC (open boxes), JLab (solid points), and CLAS (crosses). The red bands display 1$\sigma$ uncertainties for the LFHQCD EMC model. The number in parenthesis next to the experiment name in the legend is the normalization factor that multiplies all the data points, $\eta_{exp}$ in Eq. (\ref{chi}).} 
\label{fig:plt4}
\end{figure}

There are strong parallels in the expressions for $F_2^A$ in this model and the SRC-based model in Ref. \cite{Schmookler:2019nvf}. Specifically, both models have expressions for $F_2^A$ which involve "free", $Z F_2^p + N F_2^n$, and medium modified contributions. Furthermore, in both models, the medium modified contribution depends on two parameters that capture the effects of a nuclear medium on the proton and neutron, multiplied by a function that is nucleus independent (the approach of using a nuclear independent universal function was also used in Ref. \cite{Alekhin:2022tip}, which controlled off shell correlations ). Motivated by this, we determined the relationships between our $\delta r_p$ and $\delta r_n$ values and the $a_2^p$ and $a_2^n$ values given in Ref. \cite{Schmookler:2019nvf}.

\begin{equation}
    a_2^p = \frac{1}{Z}\frac{\sigma_A(Q^2,x)}{\sigma_d(Q^2,x)} \Bigg|_{Q^2 > 1.5, \,\, 1.5 \,\,  \leq \,\, x \,\, \leq 2} \,\,\, , \,\,\,\,\,\, a_2^n = \frac{1}{N}\frac{\sigma_A(Q^2,x)}{\sigma_d(Q^2,x)}\Bigg|_{Q^2 > 1.5, \,\, 1.5 \,\, \leq \,\,  x \,\, \leq 2}
\end{equation}


\begin{equation} \label{a2_relations}
    a_2^p = \frac{\delta r_p}{\delta r(^2H)}, \,\,\,\,\,\,\,\,  a_2^n = \frac{\delta r_n}{\delta r(^2H)}
\end{equation}

Where $A$ is the atomic number, $Z$ is the number of neutrons, $\sigma_A$ is the cross section of nucleus $A$, $\sigma_d$ is the cross section of deuterium, and the evaluation of $Q^2$ is in units of GeV$^2$. $a_2^p$ and $a_2^n$ are the per-proton and per-neutron SRC scaling coefficients, which are interpreted as the relative abundance of high-momentum nucleons in the measured nucleus relative to deuterium; they take into account the kinematic plateaus due to SRCs in inclusive QE scattering. With this parallel between both models in mind, our results for $\delta r_{(p,n)}$ from fitting are subject to the same relation between $a_2^p$ and $a_2^n$.

\begin{equation}
    \frac{a_2^p}{a_2^n} = \frac{\delta r_p}{\delta r_n} = \frac{N}{Z}
\end{equation}

\begin{figure}[H]
    \centering
    \includegraphics[width=5.5in]{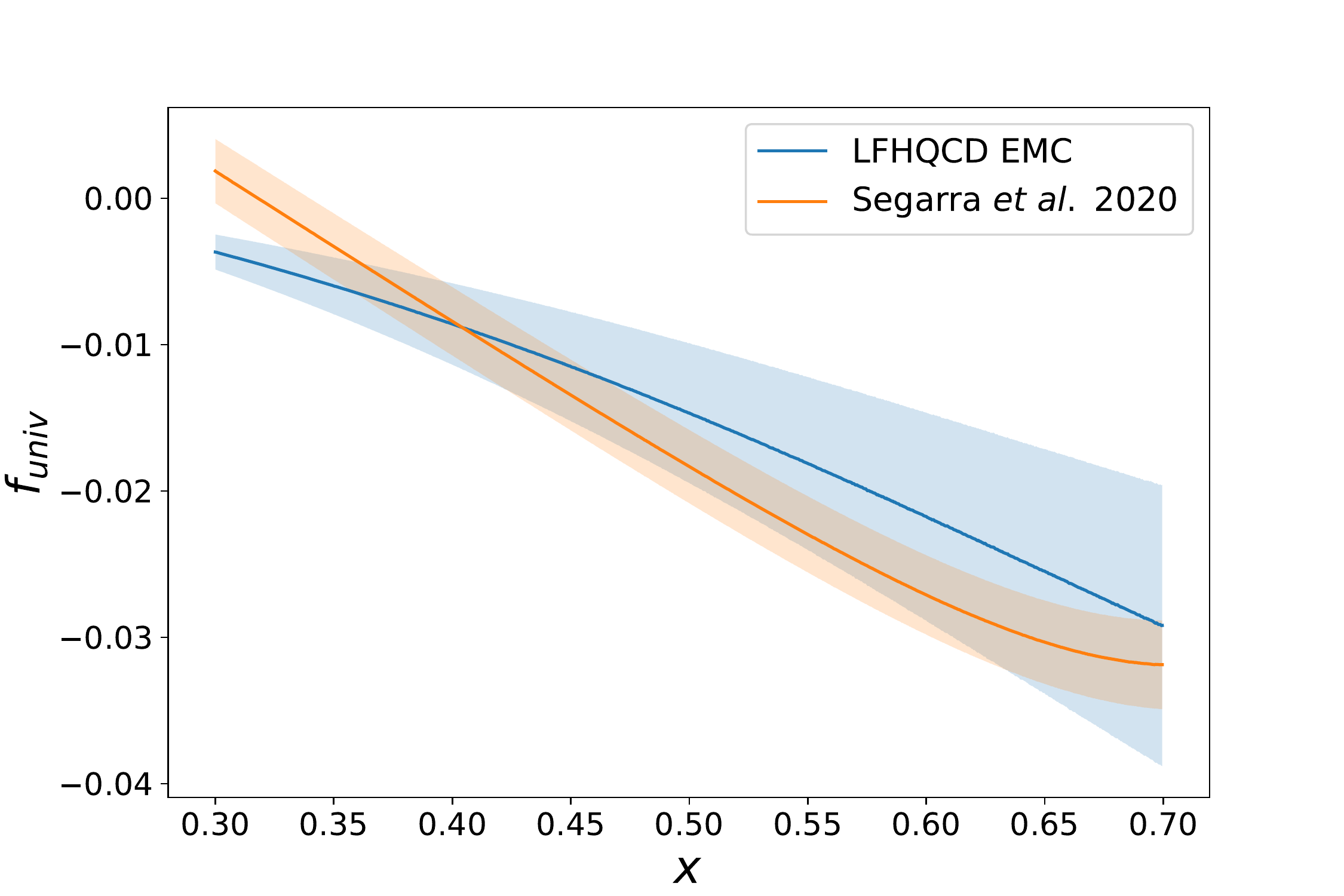}
    \caption{Comparison between Deuterium EMC ratio predicted by the LFHQCD Model (blue line) and by the universal function (orange line) parameterized in Segarra $\it{et}$ $\it{al.}$ 2020: Ref. \cite{segarra_neutron_2020}. The blue and orange shaded regions are the 1$\sigma$ uncertainty bands for the LFHQCD Model and universal function respectively.}
    \label{fig:univ_func_comp}
\end{figure}

Fig. \ref{fig:univ_func_comp} shows a comparison between the universal function in Ref. \cite{Schmookler:2019nvf}, parameterized in Ref. \cite{segarra_neutron_2020}, and the LFHQCD EMC model. They both agree within uncertainty within $0.35 < x < 0.7$.

\begin{table}[H]
    \centering
    \begin{tabular}{|P{1.5cm}|P{2.2cm}|P{2.2cm}|P{2.2cm}|P{2.2cm}|P{2.2cm}|P{2.2cm}|}
    \hline
    \multicolumn{1}{|c}{Nucleus} & \multicolumn{2}{|c|}{This Work} & \multicolumn{2}{c|}{B. Schmookler $\it{et}$ $\it{al.}$ 2019} & \multicolumn{2}{c|}{O. Hen $\it{et}$ $\it{al.}$ 2012} \\ \cline{2-7}
    \rule{0pt}{3ex}
    & $\delta r_p$ & $\delta r_n$ & $\delta r_p$ & $\delta r_n$ & $\delta r_p$ & $\delta r_n$ \\
    \hline
    \rule{0pt}{3ex} 
    $^2$H & 0.010 $\pm$ 0.003 & 0.010 $\pm$ 0.003 & & & & \\
    \hline
    \rule{0pt}{3ex} 
    $^3$He & 0.031 $\pm$ 0.003 & 0.061 $\pm$ 0.006 & & & 0.016 $\pm$ 0.005 & 0.03 $\pm$ 0.01\\
    \hline
    \rule{0pt}{3ex} 
    $^3$H & 0.032 $\pm$ 0.006 & 0.016 $\pm$ 0.003 & & & & \\
    \hline
    \rule{0pt}{3ex} 
    $^4$He & 0.040 $\pm$ 0.004 & 0.040 $\pm$ 0.004 & & & 0.04 $\pm$ 0.01 & 0.04 $\pm$ 0.01 \\
    \hline
    \rule{0pt}{3ex} 
    $^9$Be & 0.044 $\pm$ 0.004 & 0.035 $\pm$ 0.003 & & & 0.045 $\pm$ 0.014 & 0.036 $\pm$ 0.012 \\
    \hline
    \rule{0pt}{3ex} 
    $^{12}$C & 0.049 $\pm$ 0.003 & 0.049 $\pm$ 0.003 & 0.046 $\pm$ 0.015 & 0.046 $\pm$ 0.015 & 0.048 $\pm$ 0.016 & 0.048 $\pm$ 0.016 \\
    \hline
    \rule{0pt}{3ex} 
    $^{27}$Al & 0.057 $\pm$ 0.003 & 0.053 $\pm$ 0.003 & 0.051 $\pm$ 0.016 & 0.047 $\pm$ 0.015 & & \\
    \hline
    \rule{0pt}{3ex} 
    $^{56}$Fe & 0.074 $\pm$ 0.003 & 0.064 $\pm$ 0.003 & 0.053 $\pm$ 0.017 & 0.046 $\pm$ 0.015 & & \\
    \hline
    \rule{0pt}{3ex} 
    $^{63}$Cu & 0.052 $\pm$ 0.003 & 0.044 $\pm$ 0.003 & & & 0.058  $\pm$ 0.019 & 0.049  $\pm$ 0.016\\
    \hline
    \rule{0pt}{3ex} 
    $^{197}$Au & 0.072 $\pm$ 0.004 & 0.048 $\pm$ 0.003 & & & 0.065 $\pm$ 0.021 & 0.044 $\pm$ 0.014\\
    \hline
    \rule{0pt}{3ex} 
    $^{208}$Pb & 0.078 $\pm$ 0.005 & 0.051 $\pm$ 0.003 & 0.062 $\pm$ 0.020 & 0.041 $\pm$ 0.013 & & \\
    \hline
    \end{tabular}
    \caption{The $\delta r_p$ and $\delta r_n$ medium modifications used in this study and ones calculated from values in B. Schmookler $\it{et}$ $\it{al.}$ 2019: Ref. \cite{Schmookler:2019nvf} and O. Hen $\it{et}$ $\it{al.}$ 2012: Ref. \cite{hen_new_2012} using Eq. (\ref{a2_relations}). The uncertainties for $\delta r_p$ and $\delta r_n$ calculated from the references were propagated using Eq. (\ref{a2_relations}). }
    \label{tab:tabulated_delta}
\end{table}

\begin{figure}[H]
\hspace{2.5cm}\includegraphics[width=4.5in]{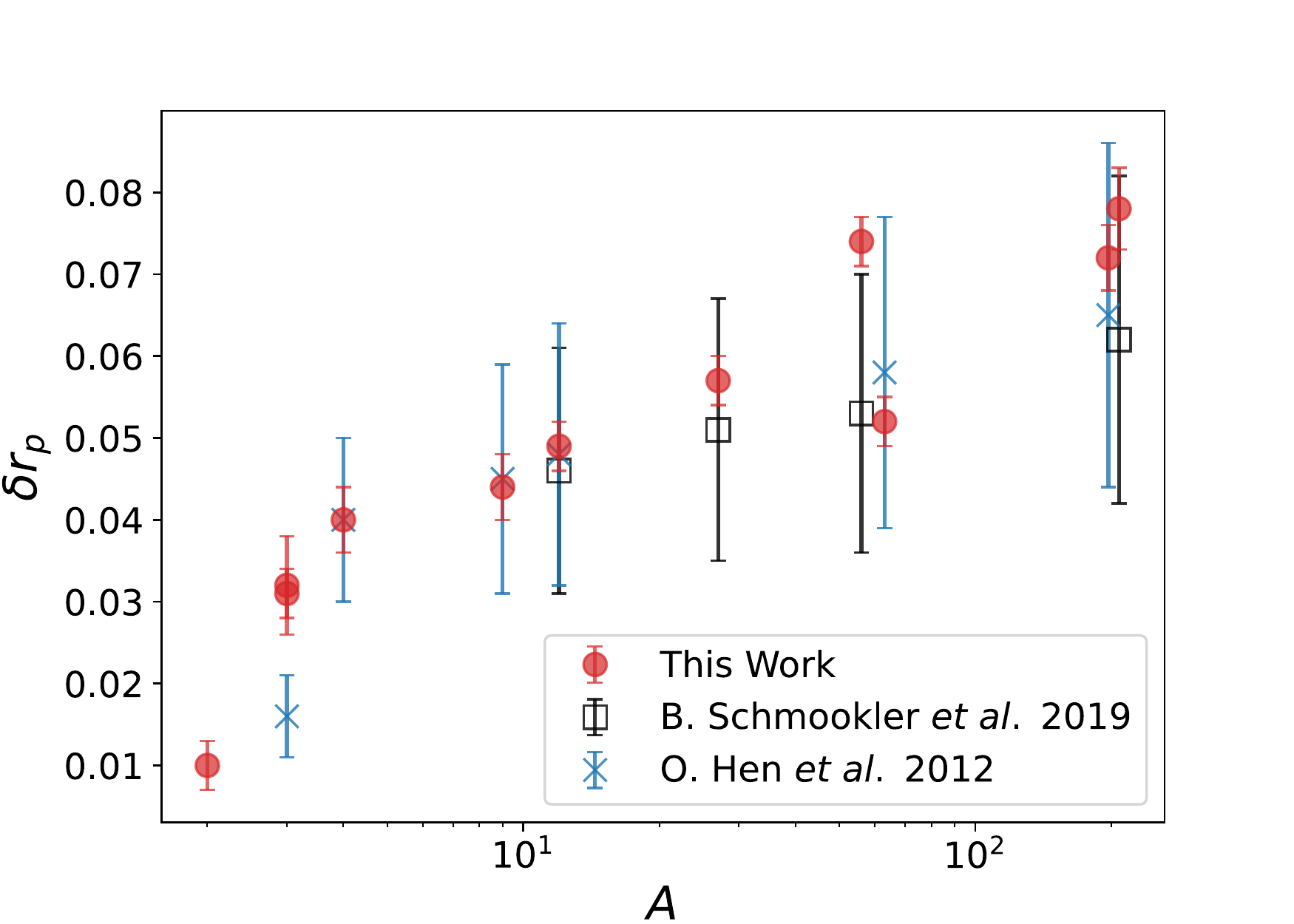}
\caption{Plotted $\delta r_p$ values as a function of mass number, A, from Table. I (B. Schmookler $\it{et}$ $\it{al.}$ 2019: Ref. \cite{Schmookler:2019nvf}, O. Hen $\it{et}$ $\it{al.}$ 2012: Ref. \cite{hen_new_2012}). } 
\label{fig:plt5}
\end{figure}

\begin{figure}[H]
\hspace{2.5cm}\includegraphics[width=4.5in]{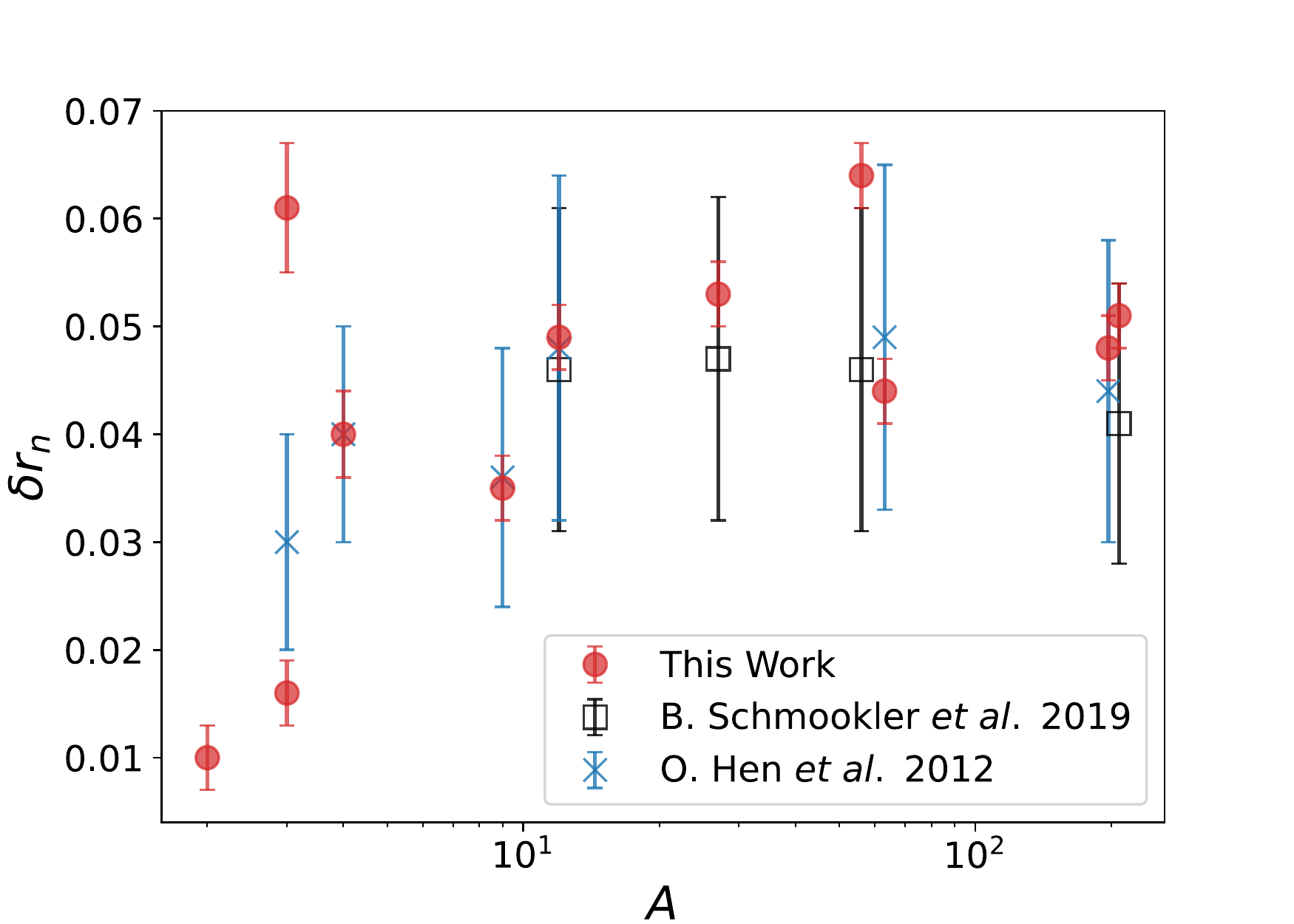}
\caption{Plotted $\delta r_n$ values as a function of mass number, A, from Table. I (B. Schmookler $\it{et}$ $\it{al.}$ 2019: Ref. \cite{Schmookler:2019nvf}, O. Hen $\it{et}$ $\it{al.}$ 2012: Ref. \cite{hen_new_2012}).} 
\label{fig:plt6}
\end{figure}

\section{Test for SRC Dominance}
From Eq. \ref{deltar_relations}, we see that $\delta r_{(n,p)}$ is proportional to the nuclear potential $|U_{(n,p)}|$, which is also proportional to the virtually. We can determine $|U_{(n,p)}|$ from our fitted $\delta r_{(p,n)}$ values by using Eq. (\ref{deltar}), and noting that the energy difference, $\bar{\Delta}$, between the nucleon ground state, $|N\ra$, and the first excited state, $|X\ra$, is approximately equal to the 500 MeV, the Roper Resonance. For $A \geq 4$, $\delta r_{(p,n)}$ values range from 0.035 to 0.078, thus $ 70 \,\, \mbox{MeV} \leq |U_{(n,p)}| \leq 156 \,\, \mbox{MeV}$. We decompose the nuclear potential into mean field and SRC contributions,

\begin{equation}
    U_{(n,p)} = U_{mean} + U_{SRC} 
\end{equation}

Using a value of 50 MeV for the absolute value of the mean field \cite{Lilley:2009zz}, obtained from the nuclear shell model, we find the model is consistent with the intuition that high virtuality is due to SRCs; by "high" virtuality, we mean virtuality greater than the typical average virtuality due to the nuclear mean field. The arguments presented here cannot be used for nuclei with $A < 4$, as the mean field is undetermined.

\section{Nucleon Charge Radius Check}
Using our values of $\delta r_{(p,n)}$, we can determine the effects of medium modifications to the charge radius. The modified electromagnetic Sachs form factor is

\begin{equation}
    \tilde{G}_{EM}^N(Q^2) = \tilde{F}_1^N(Q^2)  - \frac{Q^2}{2M^2} \tilde{F}_2^N(Q^2).
\end{equation}

Where $F_1^N$ is obtained from Eqs. (\ref{f1p_nomod}, \ref{f1n_nomod}), $\tilde{F}_1^N$ is obtained from Eqs. (\ref{f1p_mod}, \ref{f1n_mod}), and $M$ is the nucleon mass which is approximately 938 MeV for the proton and neutron. The elastic form factors $F_2^N$ from LFHQCD are obtained from Ref. \cite{sufian_analysis_2017},

\begin{equation}
    F_2^p = \chi_p [   (1- \gamma_p)F_{\tau = 4} + \gamma_p F_{\tau = 6}],
\end{equation}

\begin{equation}
    F_2^n = \chi_n [   (1- \gamma_n)F_{\tau = 4} + \gamma_n F_{\tau = 6}].
\end{equation}

\noindent Where $\chi_p = 1.793$ is the proton anomalous moment, $\chi_n = -1.913$ is the neutron anomalous moment, and $\gamma_p$ and $\gamma_n$ are the higher Fock probabilities given as 0.27 and 0.38 respectively from Ref. \cite{sufian_analysis_2017}. It is important to note that in our study, $F_2^N$ is not modified by the nuclear medium because it's expression does not involve effective charges. The change in the charge radius is given by

\begin{equation}
\sqrt{\frac{d \tilde{G}_{EM}^N (Q^2)}{d Q^2}\Bigg|_{Q^2 = 0} \Bigg/ \frac{d G_{EM}^N (Q^2)}{d Q^2}
\Bigg|_{Q^2 = 0}}
\end{equation}

Using the largest values for $\delta r_p$ and $\delta r_n$ from our fits, we find that, in the sample of nuclei that we have studied, the greatest increase in the proton and neutron charge radius is by 0.48\% and 2.9\% respectively. These results are consistent with an upper limit on the charge radius increase of 3.6\% given in Ref. \cite{mckeown_precise_1986}.






\section{Summary \& Discussion}
The results presented here provide a new model for the EMC effect using LFHQCD, motivated by a two-component model of the nucleon. The model suggests that the EMC effect is a result of the nuclear potential further breaking SU(6) symmetry. The effects of a nuclear medium are applied through two free parameters, $\delta r_p$ and $\delta r_n$, which modify the effective charges of a proton and neutron with positive and negative chiralities. The LFHQCD EMC model has strong parallels with the phenomenological model presented in Ref. \cite{Schmookler:2019nvf} in that the nuclear structure functions have contributions from uncorrelated nucleons and correlated nucleons in SRC pairs. As such, the model displays a connection with the correlation between the EMC effect and the SRC pair density. This model leads to good description of the EMC effect for a variety of nuclei, and gives results regarding changes to the proton and neutron charge radii that are consistent with Ref. \cite{mckeown_precise_1986}. A further study into the medium modification of nucleon wave functions would lead to a complete description of nuclear modifications in LFHQCD, as well as provide the first corrections to this model. Additionally a study into the nuclear potential's $A$ and $Z$ dependence would provide useful insight into the analytic form of $\delta r_{(n,p)}$. The degeneracy in the BLC and PLC due to $r$ = 3/2 leads to $\Delta = 0$, meaning that the PLC and BLC energy is degenerate, and is an unresolved issue.




\section*{Acknowledgements}

 D. N. Kim and G. A. Miller would like to thank Barak Schmookler, Or Hen, John Arrington, Guy F. de Teramond, Stanley J. Brodsky, Andrew W. Denniston, and Jackson Pybus for useful discussions. This work was supported by the U. S. Department of Energy Office of Science, Office of Nuclear Physics under Award Number DE-FG02-97ER-41014.

\bibliography{miller}{}

\begin{thebibliography}{35}
\bibitem [{pri()}]{privateconvo}%
  \BibitemOpen
  \bibfield  {title} {\enquote {\bibinfo {title} {{Results obtained from Guy F.
  de Teramond and Stanley J. Brodsky through private discussions, to be
  published in the EPJC Volume celebrating 50 years of Quantum Chromodynamics,
  edited by Franz Gross and Eberhard Klempt.}}}\ }\href@noop {} {\
  }\BibitemShut {NoStop}%
\end{thebibliography}

\end{document}